\pdfoutput=1

\documentclass[pra,10pt,superscriptaddress,footnoteinbib]{revtex4}
\usepackage{amsmath}
\usepackage{latexsym}
\usepackage{amssymb}
\usepackage{graphics,epstopdf}
\usepackage[colorlinks=true, citecolor=blue, urlcolor=blue ]{hyperref}
\usepackage{epsf,graphics,graphicx}

\newcommand{\n}{\noindent}

\newcommand{\ed}{\end{document}}
\newcommand{\beq}{\begin{equation}}
\newcommand{\eeq}{\end{equation}}

\begin{document}

\title{The effect of inertia on the Dirac electron, the spin Hall current and the momentum space Berry curvature }

\author{Debashree Chowdhury\footnote{Electronic address:{debashreephys@gmail.com}}${}^{}$  and B. Basu\footnote{Electronic
address: {sribbasu@gmail.com}} ${}^{}$} \affiliation{Physics and
Applied Mathematics Unit, Indian Statistical Institute, 203
B.T.Road, Kolkata 700 108, India}


\begin{abstract}
\n
We have studied the spin dependent force and the associated momentum space Berry curvature in an accelerating system. The results are derived by taking into consideration the non relativistic limit of a generally covariant Dirac equation under the  presence of electromagnetic field where the methodology of Foldy-Wouthuysen transformation is applied to achieve the non relativistic limit. Spin currents appear due to the combined action of the external electric field, crystal field  and the induced  inertial electric field via the total effective spin-orbit interaction. In an accelerating frame, the crucial role of momentum space Berry curvature in  the spin dynamics has also been addressed from the perspective of spin Hall conductivity. For time dependent acceleration, the expression for the spin polarization has been derived.
\end{abstract}

%



\maketitle

\section{Introduction}
In recent times, there is a growing attention in the field of $spintronics$. In this broad area of research, one studies the quantum transport properties of the electron spins and its application to technology \cite{wolf,zutic,sh1}. As spin current is non-conserved quantity, the control and generation of spin current is a challenging task. Since, the theoretical prediction of the spin Hall effect (SHE) \cite{spinh}, the application of spintronics has seen considerable advancement. This effect is observed experimentally in semiconductors and metals
 \cite{sh2}. SHE is a form of anomalous Hall effect induced by spin. Here a beam of particles separates in to up and down spin projections in the presence of perpendicular electric field in analogy to Hall effect where charges are separated in a beam passing through a perpendicular magnetic field. Besides, the spin orbit interaction opens up the possibility of manipulating electron (or hole) spin in non magnetic materials by electrical means \cite{29, 30}
and as such attracted a lot of interest of theoreticians and experimentalists recently. SHE occurs due to the spin orbit coupling(SOC) of electron with impurities is known as extrinsic, whereas the intrinsic SHE is due to SOC in band structure of semiconductor without the presence of disorder, this become an active area of research \cite{50,51,52,54,56,58, 60}. However, though the studies on the inertial effect of electrons has a long standing history \cite{barnett,ein,tol,o} but the contribution of the spin-orbit interaction (SOI) in accelerating frames has not much been addressed in the literature. Very recently, there has been an elegant attempt to extend the theory of spin current in the inertial frame \cite{matsuoprl}. A theory has been  proposed describing the direct coupling of the mechanical rotation and spin current and predicting the spin current generation arising from rotational
motion.\cite{matsuoprl,c}.

 On the other hand,  discovery of the Berry phase \cite{berry} has shed new light  in the understanding of the  origin of many physical phenomena from  an intriguing perspective . The Berry phase approach has been successfully adopted to explain various quantum mechanical/semiclassical  aspects of different systems in condensed matter physics \cite{xiaorev,bpb,carolo,zhu,bb,pra2}. In particular, the studies on momentum space Berry phase associated to spin Hall effect \cite{shen,46,78,sg}has become a topic of great recent interest . One of the avenues to investigate the motion of spins through gauge theories entails the study of the gauge fields, which naturally couples to the spin. In the field of spintronics, the theories  based on the Berry gauge theories provide a deeper understanding of the physical effects and their origin. In the semi classical equations of motion, the Berry curvature corrections have broad impacts on transport properties \cite{xiaorev}. In momentum space the Berry gauge connection due to spin-orbit coupling arise in spintronic, optical \cite{horvathy,bliokh} and graphene systems \cite{graph1,graph2}.

In a recent paper \cite{c}, the spin dependent inertial force is studied in the absence of external electromagnetic field.
In this context, it would be interesting to study the generation of spin current in an accelerating frame under the
presence of electromagnetic field and explore the conditions for the appearance of momentum space Berry curvature.
In this paper, we study the Inertial Spin Hall effect which
refers to the fact that we investigate the conditions of the  spin-dependent inertial forces and spin currents that appear in accelerating frames.
In our formulation, we consider the fundamental Dirac Hamiltonian for a particle in a linearly accelerating frame \cite{o}. As the dynamics of spin currents is
closely related to the spin-orbit interaction (SOI), we consider the low energy limit of the Dirac equation.

The formalism adopted by us is explained in detail in Section II. We have dealt with the idea \cite{c} of interpreting the effect of linear acceleration on an electron as induced effective electric field in this section. The spin-orbit interaction resulting from this induced electric field along with electric field due to the spin-orbit interaction generated from the external electromagnetic field produces the spin current in our system.  In the subsection II D we follow the physically intuitive
approach of Chudnovsky \cite{n}, based on an extension of the Drude model which accounts for
 the spin and SOI
and  derive the  spin Hall current and conductivity.
The connection of SHE and momentum  space Berry curvature motivated us to explain the physical consequences of the momentum space Berry curvature in the inertial spin Hall effect, which is the content of Section III. The paper ends with Conclusions in section IV.
\section{Dirac equation in a linearly accelerating frame}
\subsection{The Hamiltonian}
Let us start by constructing  the Dirac equation in
a non-inertial frame, following the work of Hehl and Ni\cite{o}. The essential idea in \cite{o}
is to introduce a system of orthonormal tetrad carried by the accelerating
observer. This in turn induces a
non-trivial metric and subsequently one rewrites the Dirac equation in the
observer's local frame where
normal derivatives are replaced by covariant derivatives derived from the
induced metric.
The  Dirac Hamiltonian in an arbitrary non-inertial frame with linear
acceleration and rotation for a charged particle having charge $e$ is given by \cite{o}

\begin{eqnarray}
H &=& \beta mc^{2} +  c\left(\vec{\alpha}.(\vec{p}-\frac{e\vec{A}}{c})\right)
+\frac{1}{2c}\left[(\vec{a}.\vec{r})((\vec{p}-\frac{e\vec{A}}{c}).\vec{\alpha}) +
((\vec{p}-\frac{e\vec{A}}{c}).\vec{\alpha})(\vec{a}.\vec{r})\right] \cr
&&+\beta m
(\vec{a}.\vec{r}) + eV(\vec{r}) - \vec{\Omega}.(\vec{L} + \vec{S})\label{q}
\end{eqnarray}

where $\vec{a}$ and  $\vec{\Omega}$ are respectively the linear acceleration and
rotation frequency of the observer with respect to an inertial frame.  $\vec{L}$
 and $\vec S$ are respectively  the angular momentum
($\vec{L} = \vec{r}\times \vec{p}$) and spin  of the Dirac particle. $ \vec{A} $ denotes the
vector potential. The Dirac matrices $ \beta$,  $\alpha $ and the  spin operator $\Sigma$ for 4-spinor are respectively given by
\begin{equation}\label{alfa}
\beta = \left( \begin{array}{cc}
I & 0 \\
  0 &-I
\end{array} \right),~~~~
\alpha= \left( \begin{array}{cc}
0 & \sigma \\
  \sigma & 0
\end{array} \right), ~~~
\Sigma = \frac{\hbar}{2}
\left(\begin{array}{cc}
\sigma & 0\\
0 & \sigma
 \end{array}\right)
\end{equation}
To study the only the effects of linear acceleration on the Dirac electron, we
drop the rotation term   $ \vec{\Omega}.(\vec{L}+ \vec{S}) $ and  consider the Hamiltonian as
\begin{equation}\label{ha}
H= \beta mc^{2} +  c\left(\vec{\alpha}.(\vec{p}-\frac{e\vec{A}}{c})\right)
+\frac{1}{2c}\left[(\vec{a}.\vec{r})((\vec{p}-\frac{e\vec{A}}{c}).\vec{\alpha}) +
((\vec{p}-\frac{e\vec{A}}{c}).\vec{\alpha})(\vec{a}.\vec{r})\right]
+\beta m
(\vec{a}.\vec{r}) + eV(\vec{r})
\end{equation}
 For further calculations in the low energy limit, one has to apply a series of Foldy-Wouthuysen  Transformations  (FWT) \cite{gre,m} on the Hamiltonian (\ref{ha} ).

\subsection{Foldy Wouthuysen Transformation}
The Dirac wave function is a four component spinor with the up and
down spin electron and hole components. Generically the energy gap between the
electron and hole is much larger than the energy scales associated with
condensed matter systems. One
can achieve this by block diagonalization method of the Dirac Hamiltonian
exploiting FWT \cite{m}.
The Hamiltonian (\ref{ha} ) can be divided into  block diagonal and off
diagonal
parts denoted by $ \epsilon $ and $ O $, respectively,
\begin{eqnarray}
H &=& \beta mc^{2} + O +\epsilon ,\\
 O &=& c\vec{\alpha}.((\vec{p}-\frac{e\vec{A}}{c})
+\frac{1}{2c}\left[(\vec{a}.\vec{r})((\vec{p}-\frac{e\vec{A}}{c}).\vec{\alpha}) +
((\vec{p}-\frac{e\vec{A}}{c}).\vec{\alpha})(\vec{a}.\vec{r})\right],\\
\epsilon &=& \beta m (\vec{a}.\vec{r}) + eV(\vec{r}).
\label{fwh}
\end{eqnarray}
where $\beta=\gamma_0$
and $\alpha_i=\gamma_0\gamma_i$. Applying FWT on $H$ yields,
\begin{equation}
H_{FW} = \beta \left(mc^{2}+\frac{O^{2}}{2mc^{2}}\right)+ \epsilon
-\frac{1}{8m^{2}c^{4}} \left[O ,[O,\epsilon]\right]
\end{equation}

The various terms of the Hamiltonian can be calculated \cite{c} as,
\begin{eqnarray}
 O^{2} &=& c^{2}\frac{(\vec{p}-\frac{e\vec{A}}{c})^{2}}{2m} -ce\hbar \vec{\Sigma} . \vec{B}\\~
\left[O,\epsilon\right] &=& -ice\hbar \vec{\alpha} . \vec{\nabla} V(\vec{r}) -\beta
mic\hbar (\vec{\alpha}.\vec{a})\\
\left[O,[O,\epsilon]\right] &=& ec^{2}\hbar (\vec{\nabla}.\vec{E}) + iec^{2}\hbar^{2}
\vec{\Sigma}.(\vec{\nabla}\times \vec{E})+
 2ec^{2}\hbar \vec{\Sigma}.(\vec{E}\times \vec{p})
 -\beta mc^{2}\hbar^{2}(\vec{\nabla}.\vec{a})\nonumber\\
 &&-i\beta mc^{2}\hbar^{2}
\vec{\Sigma}.(\vec{\nabla}\times \vec{a}) - 2\beta mc^{2}\hbar
\vec{\Sigma}.(\vec{a}\times\vec{p})
\end{eqnarray}
 where we have used the following relation,
\beq (\vec{\alpha} . \vec{A})(\vec{\alpha} .\vec{B}) =  \vec{A}.\vec{ B} +
i\vec{\Sigma}.(\vec{A}\times\vec{B} ).\eeq

In the above calculations we have neglected the $\frac{1}{c^{4}}$ terms and the
terms due to red shift effect of kinetic energy.
Adding all these terms the FW transformed Hamiltonian takes the form
\begin{equation}
H_{FW} = \beta\left( mc^{2} + \frac{(\vec{p}-\frac{e\vec{A}}{c})^{2}}{2m}\right) + eV(\vec{r}) + \beta m (\vec{a}.\vec{r})$$$$
-\frac{e\hbar}{2mc}\vec{\Sigma}.\vec{B}-  \frac{e\hbar^{2}}{8m^{2}c^{2}}
(\vec{\nabla}.\vec{E}) -
\frac{ie\hbar^{2}}{8m^{2}c^{2}}\vec{\Sigma}.(\vec{\nabla}\times \vec{E})-
\frac{e\hbar}{4m^{2}c^{2}}\vec{\Sigma}.(\vec{E}\times \vec{p})$$$$
+\frac{\beta\hbar^{2}}{8mc^{2}}(\vec{\nabla}.\vec{a})
+ \frac{i\beta\hbar^{2}}{8mc^{2}}\vec{\Sigma}.(\vec{\nabla}\times \vec{a}) +
\frac{\beta\hbar}{4mc^{2}}\vec{\Sigma}.(\vec{a}\times \vec{p})\label{w}
\end{equation}
Let us simplify the Hamiltonian a little bit. As we are dealing with the constant acceleration we can drop the terms $ (\vec{\nabla}.\vec{a})$ and $\vec{\Sigma}.(\vec{\nabla}\times \vec{a})$.
 Consideration of constant electric field can help us leaving  the terms with  $(\vec{\nabla}\times \vec{E})$ and
$(\vec{\nabla}.\vec{E})$. Finally, we land up with the Hamiltonian for the upper component of Dirac spinor as
\begin{equation}
H_{FW} = \left( mc^{2} + \frac{(\vec{p}-\frac{e\vec{A}}{c})^{2}}{2m}\right) + eV(\vec{r}) +  m (\vec{a}.\vec{r})
-\frac{e\hbar}{2mc}\vec{\sigma}.\vec{B}-
\frac{e\hbar}{4m^{2}c^{2}}\vec{\sigma}.(\vec{E}\times \vec{p})
+\frac{\beta\hbar}{4mc^{2}}\vec{\sigma}.(\vec{a}\times \vec{p})\label{w1}
\end{equation}
This FW transformed
Hamiltonian gives  the dynamics of an electron (or hole with proper sign of
$e$) in the positive energy part of the full energy spectrum. Here $\vec{\sigma}$ is the Pauli spin matrix.
We make a point here that the inertial effect of the linear acceleration on electron can be interpreted as an  induced $effective$ $electric$ $field$ $\vec{E}_{\vec{a}}$ \cite{c} such that \beq \vec{E}_{\vec{a}} = \frac{m}{e}\vec{a}.\eeq
Introduction of this effective electric field $\vec{E}_{\vec{a}}$ generates an inertial SOI term apart from the SOI term arising due to the external electric field (forth term in the right hand side of (\ref{w1})). The third term and the last term in the R.H.S of the Hamiltonian (\ref{w1}), explains the inertial effect due to the linear acceleration of the system. Hamiltonian (\ref{w1}) thus can be rewritten in terms of $\vec{E}_{\vec{a}}$ and potential $V_{\vec{a}} = -\vec{E}_{\vec{a}}.\vec{r}$ as  \begin{equation}\label{hfw2}
H_{FW}= \left(mc^{2} + \frac{(\vec{p}-\frac{e\vec{A}}{c})^{2}}{2m}\right) + eV(\vec{r})-\frac{e\hbar}{2mc}\vec{\sigma}.\vec{B} -  e V_{\vec{a}}(\vec{r})
-\frac{e\hbar}{4m^{2}c^{2}}\vec{\sigma}.(\vec{E}\times \vec{p})+ \frac{e\hbar}{4m^{2}c^{2}}\vec{\sigma}.(E_{\vec{a}}\times \vec{p}).
\end{equation}
We are now in a position to explain the underlying physics of the individual terms of the right hand side of the Hamiltonian (\ref{hfw2}). The first two terms describe the relativistic mass increase,  whereas the third  term is the
electrostatic energy and the fourth term is a magnetic dipole energy
which induces Zeeman effect. The term
 $\frac{e\hbar}{4m^{2}c^{2}}\vec{\sigma}.(\vec{E}\times \vec{p}) $ is the spin-orbit interaction term. The fifth term and the last term arises due to linear acceleration in the system.  It is noteworthy that the induced  electric field due to acceleration
produces another spin orbit interaction term as
 $\displaystyle {\frac{e\hbar}{4m^{2}c^{2}}\vec{\sigma}.(\vec{E}_{\vec a}\times \vec{p})}.$ The total spin-orbit interactions
 \beq \frac{e\hbar}{4m^{2}c^{2}}\vec{\sigma}.\left((\vec{E} -\vec{E}_{\vec a})\times \vec{p}\right) , \eeq
 play an effective role for our further calculations.

\subsection{Equations of motion and spin current}
Collecting the dynamical terms and the terms due to spin orbit interaction, the
final Hamiltonian for the positive energy solution of spin $ \frac{1}{2}$ electron can now be read  as
\begin{equation}
H_{FW} =  \frac{\vec{p}^{2}}{2m} + eV(\vec{r}) - eV_{a}(\vec{r})
- \frac{e\hbar}{4m^{2}c^{2}}\vec{\sigma}.(\vec{E}\times \vec{p}) +
\frac{e\hbar}{4m^{2}c^{2}}\vec{\sigma}.(\vec{E}_{\vec{a}}\times \vec{p})\label{12344}
\end{equation}

 To evaluate the semiclassical equation of motion we use the quantum mechanical analogue of force operator $\vec{F},$ defined as
\beq \vec{F} = \frac{1}{i\hbar}\left[m\vec{\dot{r}},H_{FW}\right] + m\frac{\partial\vec{\dot{r}}}{\partial t},\eeq with
$ \vec{\dot{r}}  = \frac{1}{i\hbar}[\vec{r}, H_{FW}].$ Thus we have
\beq \vec{\dot{r}} = \frac{\vec{p}}{m} -
\frac{e\hbar}{4m^{2}c^{2}}\left(\vec{\sigma}\times \vec{E})\right)
+\frac{e\hbar}{4m^{2}c^{2}}(\vec{\sigma}\times \vec{E}_{a})\label{m}\eeq
Finally, the force have the following form,
\begin{eqnarray}
\vec{F} = m\ddot{\vec{r}} = -e\vec{\nabla}V(\vec{r}) +e\vec{\nabla}V_{a}(\vec{r})
+ \frac{e\hbar}{4mc^{2}}\dot{
\vec{r}}\times\vec{\nabla}\times\left(\vec{\sigma}\times \vec{E}\right)
-\frac{e\hbar}{4mc^{2}}\dot{ \vec{r}}\times\vec{\nabla}\times
(\vec{\sigma}\times \vec{E}_{a})\label{18}
\end{eqnarray}
or,
\begin{eqnarray}\label{lor1}
\vec{F} = -e\vec{\nabla}\left(V(\vec{r})- V_{\vec{a}}(\vec{r})\right)
-\dot{\vec{r}}\times \vec{\nabla}\times
\left(\frac{e\hbar}{4mc^{2}}(\vec{\sigma}\times \vec{\nabla} V(\vec{r}))
 +\frac{e\hbar}{4mc^{2}}(\vec{\sigma}\times \vec{E}_{\vec{a}})\right)
\end{eqnarray}
It is worth mentioning here that this spin dependent effective Lorentz force noted in eqn.(\ref{lor1}), is responsible for the transport of the electrons in the system, and hence responsible for the spin Hall effect of this inertial system.
From the expression of $\dot{\vec{r}}$  in (\ref{m}) we can write the linear
velocity in linearly accelerating frame
\beq
\displaystyle {\dot{\vec{r}} = \frac{\vec{p}}{m} + \vec{v}_{\vec{\sigma},~{\vec{a}}}} \eeq where
\begin{eqnarray}\label{mnn}
\displaystyle \vec{v}_{\vec{\sigma},~\vec{a}} &=& \frac{e\hbar}{4m^{2}c^{2}} \vec{\sigma}\times\left[\vec{\nabla}
V(\vec{r}) - \vec{\nabla}
V_{\vec{a}}(\vec{r})  \right] \nonumber \\
&=&
-\frac{e\hbar}{4m^{2}c^{2}} \vec{\sigma}\times\vec{ E}_{tot}
\end{eqnarray}
is the effective spin dependent velocity with
\beq\vec{ E}_{tot} =-\left [\vec{\nabla} V(\vec{r}) - \vec{\nabla}
V_{\vec{a}}(\vec{r})\right] \eeq
being the $total$ $effective$ $electric$ $field$ present in the system. One should note here that the velocity term in (\ref{mnn}) is dependent on the potential $\vec{\nabla} V(\vec{r})$ and $\vec{\nabla}V_{\vec{a}}(\vec{r}).$
Thus the inertial effect on linear acceleration produces the anomalous velocity term which in turn may yield the spin Hall effect.

 The polarized spin current  due to the total effective electric field $
\displaystyle \vec{E}_{tot} $ is  thus given by
\beq j^{i}_{s} = e~n~Tr\sigma_{i}\vec{v}_{\vec{\sigma},~\vec{a}}.\eeq
Thus the $i^{th}$ component of the spin current is given by
\beq |j^{i}_{s}| = \frac{n~e^{2}\hbar}{2m^{2}c^{2}}(\vec{S}\times \vec{E}_{tot})^{i}\eeq
where $\vec{S}$ is the spin vector.
This result is consistent with the result given in \cite{c}, where the authors have considered the induced electric field due to acceleration, but the system is not in the presence of any external electromagnetic field. With a careful observation of the force equation one can put further analysis on different components of the total current produced in an
accelerating system by adopting the averaging methodology followed in \cite{n}.
\subsection{Spin Hall conductivities}
The next job is to evaluate the explicit expression for the spin Hall conductivity in an accelerated system. To proceed in this direction we start from equation(\ref{18}) as
\begin{eqnarray}\label{lor}
\vec{F} = \vec{F}_{0}+ \vec{F}_{\vec{\sigma}}
\end{eqnarray}
where $ \vec{F}_{0}$ is the  spin independent part of the force and $\vec{F}_{\vec{\sigma}}$ is the spin dependent part of the total spin force.
The explicit forms of the above mentioned terms are
\begin{eqnarray}\label{sp0}
\vec{F}_0 & =& -e\vec{\nabla}V(\vec{r})+e\vec{\nabla}V_{\vec{a}}(r)=-e\vec{\nabla}V_{tot},\\
\vec{F}_{\vec{\sigma}} &=& -\frac{e\hbar}{4mc^{2}}\dot{ \vec{r}}\times
(\vec{\nabla}\times(\vec{\sigma}\times \vec{\nabla} V(\vec{r}))
 + \frac{e\hbar}{4mc^{2}}\dot{\vec{r}}\times(\vec{\nabla}\times(\vec{\sigma}\times\vec{\nabla}V_{a}))\label{s}
\end{eqnarray}

Here we have neglected $1/c^{4}$ terms.
One can write the total Lorentz force as
\beq \vec{F}_0 + \vec{F}_{\vec{\sigma}} = e\vec{E}_{tot} + \frac{e}{c}(\dot{\vec{r}} \times\vec{ B}(\vec{\sigma}))\label{sigma},\eeq where the spin Lorentz force is
\beq  \vec{F}_{\vec{\sigma}} =  \frac{e}{c}(\dot{\vec{r}} \times \vec{B}(\vec{\sigma}))\label{sigma1}.\eeq
This spin dependent Lorentz force is responsible for the spin current produced in the system. $ \vec{B}(\vec{\sigma})$, the $effective$ $magnetic$ $field$ appearing in the spin space can be read as
\begin{eqnarray}\label{asigma}
\vec{B}(\vec{\sigma}) = \vec{\nabla}\times \vec{A}(\vec{\sigma})
\end{eqnarray}
where the forms of the vector potentials are explicitly given by
\begin{eqnarray}
\vec{A}(\vec{\sigma}) =   -\frac{\hbar}{4mc}(\vec{\sigma}\times \vec{\nabla} V(\vec{r}))
 + \frac{\hbar}{4mc}(\vec{\sigma}\times \vec{\nabla} V_{\vec{a}}(\vec{r}))
\end{eqnarray}
Finally, the Hamiltonian (\ref{12344}) can be written as
\beq H_{FW} = \frac{1}{2m}(\vec{p} -\frac{e}{c}\vec{A}(\vec{\sigma}))^{2} + eV_{tot}(\vec{r}) \eeq
where $ V_{tot}(\vec{r}) = V(\vec{r}) - V_{\vec{a}}(\vec{r}).$ Let us remind here that the space dependent potential $V(\vec{r})$  is the sum
of the external electric potential $ V_{0}(\vec{r})$ and the lattice electric
potential $ V_{l}(\vec{r}) $.
From eqn. (\ref{lor1}), it is observed that the contribution from $\vec{ F}( \vec{\sigma})$ in comparison to $ \vec{F}_{0}$ is very small. We treat this  as a perturbation \cite{n} in the next part of our calculation. Breaking into different parts, the solution of
equation (\ref{18}) can be written as $\dot{ \vec{r}} = \dot{ \vec{r}}_{0} +
\dot{ \vec{r}}_{\vec{\sigma}}. $ If the relaxation time $\tau$ is independent of $\vec{\sigma}$ and  for the constant total electric field  $\vec{E}_{tot}$, following \cite{n} we can write,
\beq \langle\dot{ \vec{r}}_{0}\rangle = -\frac{\tau}{m}\left\langle\frac{\partial V_{tot}}{\partial r}\right\rangle =
\frac{e\tau}{m}\vec{E}_{eff}, \label{r0dot} \eeq
where we denote  $\vec{E}_{eff}=-e \vec{\nabla}\left(V_{0}(\vec{r}) - V_{\vec{a}}(\vec{r})\right).$

\begin{equation}
~~~~~~~~~~~~~~~~~\left\langle\dot{\vec{r}}(\vec{\sigma})\right\rangle = - \frac{\hbar
e^{2}\tau^{2}}{4m^{3}c^{2}}\vec{E}_{eff}\times\left\langle\frac{\partial }{\partial r} \times (\vec{\sigma}
\times\frac{\partial V_{l}}{\partial r} )\right\rangle + \frac{\hbar\tau^{2}e^{2}}{4m^{3}c^{2}} \vec{E}_{eff}\times\left\langle\frac{\partial }{\partial r} \times
(\vec{\sigma} \times \frac{\partial V_{\vec{a}}}{\partial r})\right\rangle\label{30}\\
\end{equation}
In the above expression of $\left\langle\dot{\vec{r}}(\vec{\sigma})\right\rangle,$ a term is present which represents the volume average of electrostatic potential $\partial_{i}\partial_{j}V_{l}(r)$. For a
cubic lattice, the only contribution permitted by symmetry is
\beq \left\langle\frac{\partial^{2}V_{l}}{\partial r_{i}\partial r_{j}}\right\rangle = \mu \delta_{ij},\label{sym}\eeq
where $\mu$ is a constant \cite{n} depending on the system.  Thus we have,
\beq \left\langle\dot{\vec{r}}(\vec{\sigma})\right\rangle = \frac{\hbar
e^{2}\tau^{2} \mu }{4m^{3}c^{2}}(\vec{\sigma}\times \vec{E}_{eff})
\label{r1dot}.\eeq
One may note that the second term in the right hand side of (\ref{30}) vanishes for constant
acceleration and so contribution from that term is zero in (\ref{r1dot}). Interestingly, effect of this term will contribute when the system is under non linear acceleration but that case is not studied here.

To calculate the spin current we now introduce the spin polarization vector $\vec{\lambda}=\langle \vec{\sigma}\rangle.$
The density  matrix of the charge carriers in spin space can be written as
\beq\rho~^{s} = \frac{1}{2}\rho(1 + \vec{\lambda}.\vec{\sigma})\label{deb1}\eeq
where $\rho$ is the total charge concentration. Using eqn (\ref{deb1}) and the eqns(\ref{r0dot}), (\ref{r1dot}) the total spin current can be derived as
\beq \vec{j} = e\left\langle\rho^{s}\vec{\dot{r}}\right\rangle = \vec{j}^{o ,\vec{a}} + \vec{j}^{s,\vec{a}}(\vec{\sigma}) \eeq
The charge component of this current in our accelerated system is
\begin{eqnarray}\label{jo}
\vec{j}^{o , \vec{a}} &=& \frac{e^{2}\tau \rho}{m}\vec{E}_{0} - \frac{e^{2}\tau \rho}{m}\vec{E}_{\vec{a}} \nonumber \\
&=& \frac{e^{2}\tau \rho}{m} (\vec{E}_{0}-\vec{E}_{\vec{a}})
\end{eqnarray}
whereas the spin component of the current in the accelerated system is given by
\begin{eqnarray}\label{js}
\vec{j}^{s, \vec{a}}(\vec{\sigma}) & =& \left(\frac{\hbar
e^{3}\tau^{2}\rho\mu}{2m^{3}c^{2}}\right)(\vec{\lambda}\times \vec{E}_{0}) - \left(\frac{\hbar
e^{3}\tau^{2}\rho\mu}{2m^{3}c^{2}}\right)(\vec{\lambda}\times \vec{E}_{\vec{a}}) \nonumber \\
&=& \left(\frac{\hbar e^{3}\tau^{2}\rho\mu}{2m^{3}c^{2}}\right)\left(\vec{\lambda}\times(\vec{E}_{0}-\vec{E}_{\vec{a}})\right)
\end{eqnarray}
The effect of constant linear acceleration  is observed in the expressions of charge and spin current in eqns.(\ref{jo}) and (\ref{js}).
We know that $\vec{E}_{\vec{a}} = \frac{m\vec{a}}{e},$ so in the absence of acceleration
$ \vec{E}_{\vec{a}}$ the expression for the spin current is consistent with that obtained in \cite{n}. One more important conclusion can be drawn from the above expression of currents $\vec{j}^{o , \vec{a}}$ and $\vec{j}^{s, \vec{a}}(\vec{\sigma})$: there exists a critical value of $\vec{a}_c,$ below which we get total current in the same direction as $\vec{E}_{0},$ whereas  for larger value of $\vec{a}$  the current flows in the opposite direction of $\vec{E}_{0}.$ The critical value of $\vec{a}_c$ can be tuned as
\beq \vec{a}_c = \frac{e}{m}\vec{E}_{0}.\eeq At this critical value of acceleration, which is practically a very large value, the total current of the system becomes zero. As an example if $E_0=(0,0,E_z)$, $a_c =\frac{e}{m}E_z \sim {\cal{O}}(10^{11})$, which happens to be a very large value.

 From the expressions of the currents (\ref{jo}), (\ref{js}), the corresponding Hall conductivities in an linearly  accelerating frame turns out to be
\begin{eqnarray}\label{mdm}
\sigma^{\vec{a}}_{H} &=& \frac{e^{2}\tau \rho}{m}\nonumber\\
\sigma^{s, \vec{a}}_{H} &=& \frac{\hbar e^{3}\tau^{2}\rho\mu}{2m^{3}c^{2}}
\end{eqnarray}
One can find out the ratio of charge and spin Hall conductivity from (\ref{mdm}),
\beq \frac{\sigma^{s , \vec{a}}_{H}}{\sigma^{\vec{a}}_{H}} = \frac{\hbar e\tau\mu}{2m^{2}c^{2}}.
\label{ratio}\eeq
As expected this ratio is the same as observed in a non accelerating system \cite{n}, $i.e.$ the acceleration of the system does not affect the ratio of spin to charge conductivity. The ratio (\ref{ratio}) only depends on the metal used. \\
As an example if we consider an $Al$ strip (which is a cubic lattice) of $25$ nm thickness, ${\vec E_0}=(0,0, E_z)$ and ${\vec a}=(0,0, a_z),$ the expression for the spin current can be written as
\begin{eqnarray}\label{jsd}
 j_{x}^{s, \vec{a}}(\vec{\sigma})  &=& (2.7\pm 0.6)\times 10^{3}\left(\lambda_{y}(E_{z} - E_{\vec{a}, z})\right),\\
 j_{y}^{s, \vec{a}}(\vec{\sigma})  &=& -(2.7\pm 0.6)\times 10^{3}\left(\lambda_{x}(E_{z} - E_{\vec{a}, z})\right),
\end{eqnarray} where the conductivity  is evaluated in the unit of $ \Omega^{-1} m^{-1}.$

\section{Inertial spin Hall effect and Berry curvature}

The fact, that in a  generic spin-orbit system there is a close connection of spin Hall conductivity and momentum space Berry curvature \cite{winkler} inspired us to study the Berry gauge and Berry curvature associated with our inertial spin-orbit Hamiltonian. The inertial spin-orbit Hamiltonian of the accelerated system is used to understand the underlying physics of the carrier dynamics of this system in the presence of Berry's curvature in momentum space. In this section we consider the Dirac Hamiltonian in a linearly accelerating frame without the electromagnetic field as derived in \cite{o}. Neglecting the rest mass energy and the red-shift effect of the kinetic energy, in the low energy limit  the Hamiltonian in a linearly accelerating frame is given by \cite{o},
\begin{equation}
H= \frac{\vec{p}^{2}}{2m} + m\vec{a}.\vec{r} +\frac{\hbar}{4mc^{2}}\vec{\sigma}.(\vec{a}\times \vec{p})
\end{equation}
In terms of the $induced$ $effective$ $field$ $\vec{E}_{\vec{a}}$ and the corresponding induced potential $V_{\vec{a}}$ the spin-orbit Hamiltonian is obtained as
 \beq H = \frac{\hbar^{2}\vec{k}^2}{2m} + V_{\vec{a}}(\vec{r}) - \frac{e\hbar^{2}}{4m^{2}c^{2}}\vec{\sigma}.(\vec{k} \times \vec{E}_{\vec{a}})),\label{kspace} \eeq
  where $\vec{p} = \hbar \vec{k}$ is the momentum, $m$ is the electrons mass.

\subsection{Momentum space Berry curvature in an accelerated frame}
The inertial effects of linear acceleration on the spin-orbit interaction can now be analyzed by the same framework as the standard Hamiltonian with SOI. The special relativity arguments can qualitatively explain the effect of SOI.
For electrons moving through a lattice, the electric field $\vec{E}$  is Lorentz transformed to an effective magnetic field $(\vec{k}\times \vec{E})\approx \vec{B}(\vec{k})$ in the rest frame of the electron. Thus from (\ref{kspace}) in a linearly accelerating system we can write
\beq H = \frac{\hbar^{2}\vec{k}^2}{2m} + V_{\vec{a}}(\vec{r}) - \gamma \vec{\sigma}.\vec{B}_{\vec{a}}(\vec{k})\label{kha},\eeq
where $\gamma$ is the coupling strength and the subscript $\vec{a}$  in $ \vec{B}_{\vec{a}}$ reminds us of the accelerated frame.

For each value of $\vec{k}$, the spin degeneracy of electrons split into two subbands $|\pm\rangle.$ The energy eigenvalues of these states are given by
 \beq \epsilon_{\pm} =  \frac{\hbar^{2}\vec{k}^2}{2m} + V_{\vec{a}}(\vec{r}) \pm \gamma |\vec{B}_{\vec{a}}(\vec{k})|.\label{epsi}\eeq
 In (\ref{epsi}), $\epsilon_{+}$($\epsilon_{-}$) denotes the energy value  when electron spin is in opposite (same) direction with $\vec{B}_{\vec{a}}(\vec{k}),$ for each $\vec{k}.$

 One can now find the Berry curvature utilizing  the method of local unitary transformations. At every point in the $\vec{k}$ space, the spin axis may be locked along the effective magnetic field direction through a local unitary transformation. In laboratory frame the magnetic field axis rotates as one moves from one point to other. A local transformation may rotate the frame in such a way that the spin axis(along z axis) points along  the unit vector $ n_{a}(\vec{k}),$ where \beq n_{a}(\vec{k}) = \frac{\vec{B}_{\vec{a}}(\vec{k})}{|\vec{B}_{\vec{a}}(\vec{k})|}.\eeq Here again the subscript '$\vec{a}$' reminds us of the accelerated frame. To rotate the laboratory frame let us choose a unitary matrix $ U = U(\vec{k}),$ which satisfies
\beq U(\vec{k})(\vec{\sigma}.\vec{n})U^{+}(\vec{k}) = \sigma^{z} .\eeq
The unitary matrix   $ U = U(\vec{k})$, rotates the reference spin axis along $\vec{B}_{\vec{a}}(\vec{k}). $ The transformed Hamiltonian is
\beq  H = U (\vec{k})H U^{\dagger}(\vec{k}) =  \frac{\hbar^{2}\vec{k}^2}{2m}  - \gamma \sigma^{z}|\vec{B}_{\vec{a}}(\vec{k})| + U(\vec{k})V_{\vec{a}}(\vec{r})U^{\dagger}(\vec{k})\label{kha1}\eeq
The potential term transforms nontrivially as $ U $ depends on $\vec{k}$ and the $\vec{k}$ space derivatives are given as $ \vec{r} = i\partial_{k}.$
It is noteworthy that in our formulation
\beq U(\vec{k})V_{\vec{a}}(\vec{r})U^{\dagger}(\vec{k}) = e\vec{E}_{\vec{a}}.\left(i\nabla_{k} + iU\nabla_{k}U^{\dagger}\right)\label{uni},\eeq
where the electric potential due to the accelerating frame is \beq V_{\vec{a}}(\vec{r}) = e\vec{E}_{\vec{a}}.\vec{r}.\eeq
Under the local transformation $U(\vec{k}),$ the position operator transforms into a covariant form as $ \vec{r} \rightarrow \vec{r} - A(\vec{k})$ where $A(\vec{k}) = -iU(k)\nabla_{k}U^{\dagger}(k)$ is a $2\times 2$ gauge field in momentum  space.

For the adiabatic transport, we can neglect the mixing of the bands
$|+\rangle$ and $|-\rangle$, and applying an Abelian approximation we can write  the momentum space Berry gauge field associated with this accelerating frame as
\beq \displaystyle{A}^{ad}_{\pm}(\vec{k}) \label{adia}= \pm\frac{1}{2}(1-cos\theta(\vec{k}))\nabla_{k}\phi(\vec{k}) \eeq
 where $\pm$ is for two bands and  $(\theta,\phi)$ are the $\vec{k}$ dependent spherical angles parameterizing the direction of $\vec{B}_{\vec{a}}(\vec{k})$ $i.e$ $cos\theta = \frac{B_{z}(\vec{k})}{|\vec{B}(\vec{k})|}$ and $tan\phi = \frac{B_{y}(\vec{k})}{\vec{B}_{x}(\vec{k})}$.

 The corresponding Berry curvature in the momentum space can then be written as
 \beq
 \Omega_k(\vec{k}) = \partial_{k_{i}} A^{ad}_{k_{j}}(\vec{k}) - \partial_{k_{j}} A^{ad}_{k_{i}}(\vec{k})
 \eeq
 which accounts for the real physical effect. Following the analogous situation of the real space Berry curvature one can write \cite{fujita}
  \beq \Omega_{k}(\vec{k},\pm) = \pm\frac{1}{2}\vec{n}_{a}(\vec{k}).\left(\frac{\partial\vec{n}_{a}(\vec{k})}{\partial k_{i}}\times \frac{\partial\vec{n}_{a}(\vec{k})}{\partial k_{j}} \right) \label{omega}\eeq
 $ \Omega(\vec{k})$ represents an effective magnetic field in the momentum space and is responsible for the orbital motion of carriers \cite{niu}.
The exact configuration of $\vec{B}(\vec{k})$ shows the explicit nature  of the curvature $\Omega_{k}(\vec{k}).$  For a particular choice  $ \vec{a}= (0, 0, a_{z})$, $ \vec{E}_{\vec{a}} = (0,0,\vec{E}_{z}),$  we can write  $\vec{B}_{a}(\vec{k}) = (a_{z}k_{y}, -a_{z}k_{x}).$
From (\ref{kha}), the SOC Hamiltonian for this accelerating frame can be written as
\beq \label{rasba}
H = \frac{\hbar^{2}\vec{k}^2}{2m} + \alpha(k_{x}\sigma_{y} - k_{y}\sigma_{x}) ,
\eeq where $\alpha $ contains the effect of $a_{z}.$  In the expression (\ref{rasba}), the coefficient $\alpha$ and the form of spin-orbit coupling are the same as Rashba coupling  \cite{rashba}. However, in the usual Rashba coupling, structural inversion asymmetry is responsible for the generation of internal field, whereas in our formulation, the electric field is induced  due to the inertial effect of acceleration.  In the accelerated frame, the effective magnetic field is directed along $\theta = \frac{\pi}{2}$ and $\phi = tan^{-1}(\frac{-k_{x}}{k_{y}})$ in momentum space. With this choice, from (\ref{adia}) we find the Berry gauge connection as,\beq {A}^{ad}_{\pm}(\vec{k}) = \pm\left(\frac{1}{2k^{2}}\right)(-k_{y}, k_{x}, 0). \eeq This gives a trivial result, as the Berry curvature is zero at all points in $\vec{k}$ space except at $\vec{k} = 0.$  The explicit form of the Berry curvature is \beq \Omega(\vec{k},\pm) = \left(0, 0, \pm\pi\delta^{2}(\vec{k})\right),\eeq which may be interpreted as an effective magnetic field in the system. The expression shows if the spin Lorentz force is present it will be active only in the $z-$ direction and the out-of-plane transverse spin current cannot be explained \cite{fujita} in this case with constant acceleration. For a complete explanation of this out of plane spin polarization which is an important aspect of spin Hall effect, we take resort to the case of time dependent acceleration.

\subsection{Momentum space Berry curvature for time dependent accelerated system}
In this subsection, we deal with the spin-orbit Hamiltonian with a time dependent acceleration $\vec{a}(t)$. As an example of time dependent acceleration we consider \cite{c}
\beq
\vec{a} = uw_{a}^{2}exp(iw_{\vec{a}}t)\vec{e}_{x} ,\label{atime}
\eeq where acceleration is induced by the harmonic oscillation with frequency $w_{\vec{a}}$ and amplitude $u.$ In this case, due to the time dependence of the acceleration $\vec{a},$ the induced effective electric field  $\vec{E}_{\vec{a}}$ is time dependent with an expression \beq\vec{E}_{\vec{a}} = \frac{muw^{2}_{a}}{e}exp(iw_{\vec{a}}t)\vec{e}_{x} \label{eatime}.\eeq As in the earlier situation with constant acceleration, we can write the time dependent Hamiltonian (\ref{kha})  as
\beq  H(t) = \frac{\hbar^{2}\vec{k}^2}{2m} + V_{\vec{a}}(\vec{r},t) -\gamma \vec{\sigma}.\vec{B}_{a}(\vec{k}(t)).\label{time} \eeq
Due to the acceleration of the carriers, time dependence of the spin orbit Hamiltonian will generate an additional component
\cite{fujita} of the effective magnetic field $\vec{B}_{\perp} = (\vec{\dot{n}}_{\vec{a}}\times \vec{n}_{\vec{a}}),$ in addition to the effective magnetic field $\vec{B}_{a}(\vec{k}).$ Here $\vec{n}_{\vec{a}} = p^{-1}(p_{y},-p_{x},0),$ is the unit vector along $\vec{B}_{a}(\vec{k}).$ This additional component basically  explains the out of plane spin polarization. Let the effective electric field due to acceleration be generated in the $x$ direction such that $\vec{E}_{\vec{a}} = E_{\vec{a},x}\hat{x}$ and  $\vec{\dot{n}}_{a} = p^{-1}(0,e\vec{E}_{\vec{a},x},0).$ As $\vec{B}_{\vec{a}}(\vec{k})$  is in the x-y plane, the term $\vec{B}_{\perp}$ completely represents an effective out of plane magnetic field component along $z$ direction. The classical spin vector in the $z$ direction can be expressed in terms of the total magnetic field $\vec{B}_{\Sigma}$, a sum total of $\vec{B}_{\vec{a}}(\vec{k})$ and $\vec{B}_{\perp},$ as \beq s_{z} = \pm \frac{1}{|\vec{B}_{\Sigma}|}\frac{\hbar}{2}(\vec{\dot{n}}_{\vec{a}}\times \vec{n}_{\vec{a}}).\hat{z} .\eeq
where $\pm$ represents the spin aligned parallel (antiparallel) to $\vec{B}_{\Sigma}.$ In the adiabatic condition, $\vec{B}_{\vec{a}}(\vec{k})\gg \vec{B}_{\perp}$ $i.e$ the electrons are mostly aligned along $\vec{B}_{\vec{a}}(\vec{k})$ with a small portion of them being aligned along $\vec{B}_{\perp}$ \cite{fujita}. In the adiabatic limit, $\vec{B}_{\Sigma}$ approaches $\vec{B}_{\vec{a}}(\vec{k})$ and  the out of plane spin polarization can be derived as
\begin{eqnarray}\label{szt}
s_{z}\approx &\pm& \frac{1}{|\vec{B}_{\vec{a}}(\vec{k})|}\frac{\hbar}{2}(\vec{\dot{n}}_{\vec{a}}\times \vec{n}_{\vec{a}}).\hat{z}\\
&=& \pm\frac{\hbar^{2}}{2\alpha p}\frac{\hbar}{2}\left(-\frac{1}{p^{2}}eE_{a ,x}p_{y}\right)\\
&=& \mp \frac{e\hbar^{3}p_{y}E_{a ,x}}{4\alpha p^{3}}.
\end{eqnarray}
Replacing the value of $E_{a ,x}$ from (\ref{eatime}) in (\ref{szt}), we obtain \beq s_{z} = \mp \frac{\hbar^{3}p_{y}muw^{2}_{a}}{4\alpha p^{3}}exp(iw_{\vec{a}}t).\eeq which gives the absolute value of $s_{z}$ as \beq |s_{z}| = \mp \frac{\hbar^{3}p_{y}muw^{2}_{a}}{4\alpha p^{3}}.\eeq We plot the variation of $s_z$ with $\omega_{\vec a}$ and $\frac{p_y}{p^3}$ keeping the amplitude $u$ fixed at 10nm in the left panel of Fig. 1. In the right panel we plot the variation of $s_z$ with $u$ and $\frac{p_y}{p^3}$ keeping the frequency $\omega_{\vec a}$ fixed at 10GHz. In the figure the constant $A=\frac{4\alpha}{m\hbar^3}.$
\n
\begin{figure}
\includegraphics[width=6.0 cm,angle=270]{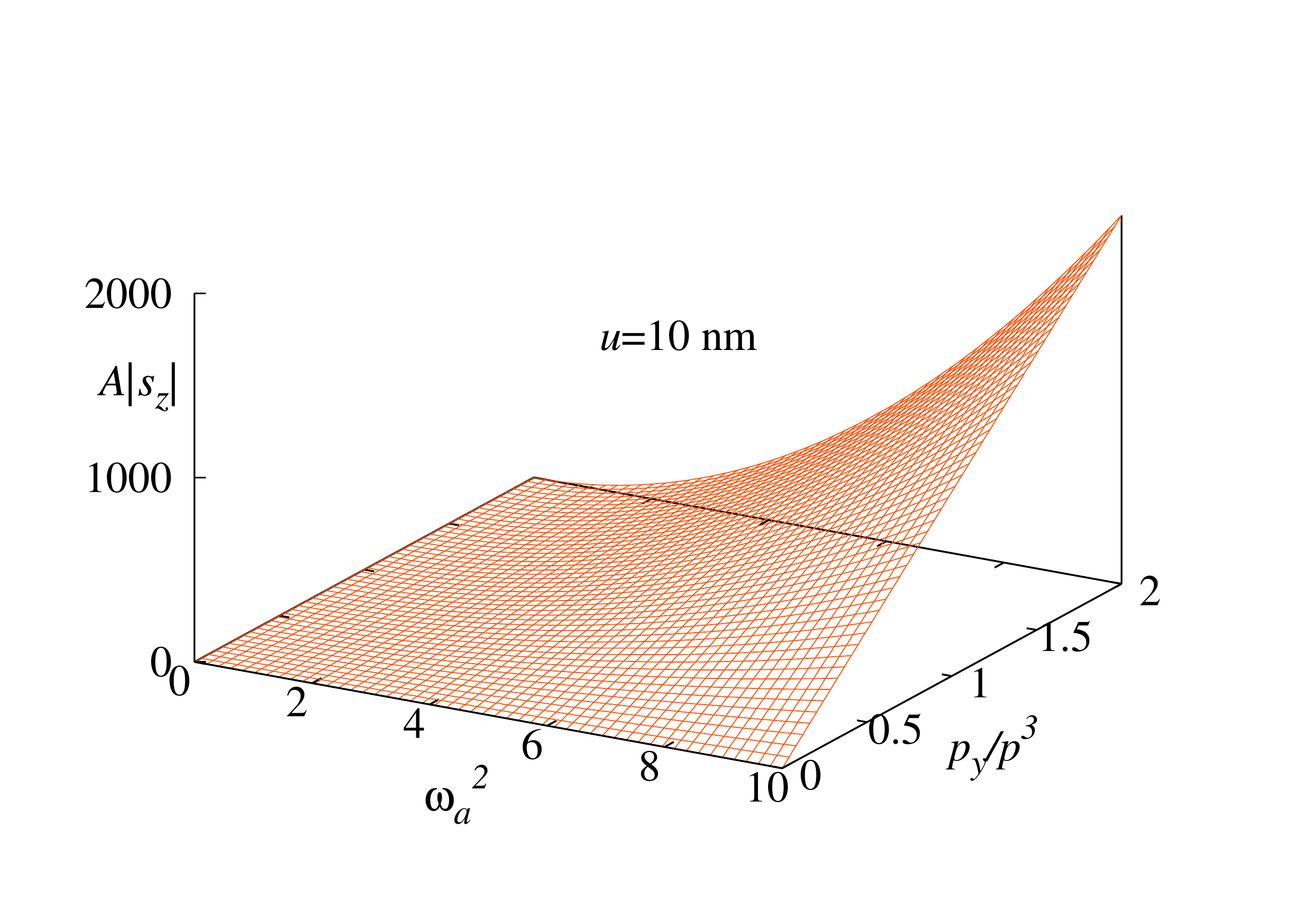}
\hspace*{-1 cm}
\includegraphics[width=6.0 cm,angle=270]{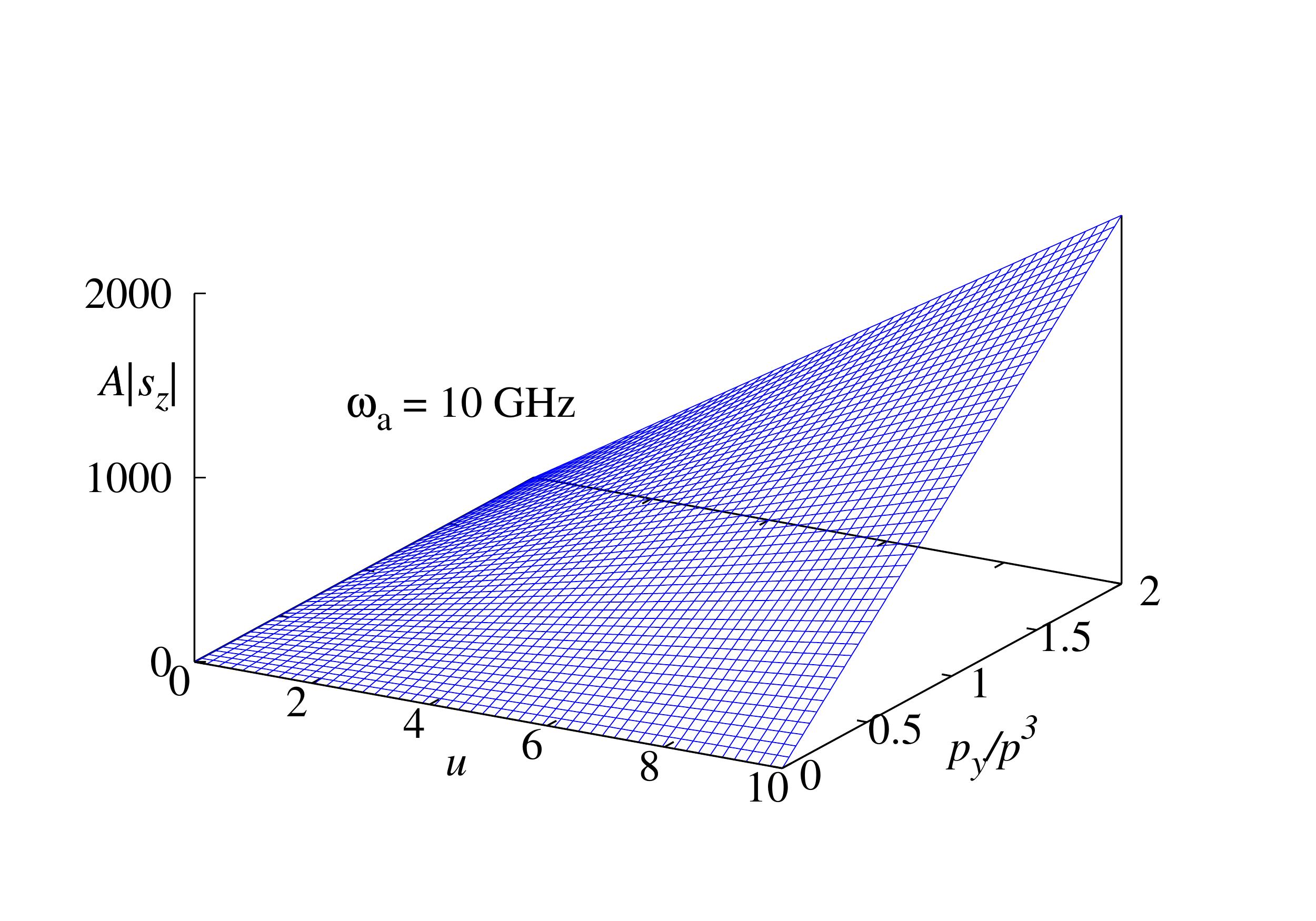}
\caption{\label{a} (Colour online) Left: Variation of $|s_z|$ with $\omega_a$ and $\frac{p_y}{p^3}$ for $u=10$ nm. Right: Variation of $|s_z|$ with $u$ and $\frac{p_y}{p^3}$ for $\omega_{a}=10$ GHz.}
\end{figure}

Another important feature of the spin transport, the spin dependent velocity  of this time dependent accelerating system can now be addressed.  Considering the
Hamiltonian (\ref{time}), the velocity along the $i$-th direction can be derived using the standard method as
\beq v_{i} = \frac{1}{i\hbar}[r_{i}, H] = \frac{p_{i}}{m} - \gamma\frac{\partial \vec{B}_{a}(\vec{k}(t))}{\partial p_{i}}.\vec{\sigma}\label{vi} \eeq
In this expression, the first term gives  the kinetic velocity which is not of our interest. We focus only on  the second term.
We write $\vec{B}_{a}(\vec{k}(t)) = |\vec{B}_{a}(\vec{k})|\vec{n}_{a}(\vec{k}(t)), $ such that the important part  of the  second term in (\ref{vi}) can be written as
\beq \frac{\partial \vec{B}_{a}(\vec{k}(t))}{\partial p_{i}} = \frac{\partial|\vec{B}_{a}(\vec{k})|}{\partial p_{i}}\vec{n}_{a} + |\vec{B}_{a}(\vec{k})|\frac{\partial\vec{n}_{a}}{\partial p_{i}} \label{bkt}\eeq
  As a result of the time dependence of the acceleration, the spin experiences  an  extra magnetic field $\vec{B}_{\perp} $\cite{fujita}
such that the system is associated with a total magnetic field $\vec{B}_{\Sigma}.$ The spin vector $\sigma$ can be written as $\sigma = \frac{\vec{B}_{\Sigma}}{|\vec{B}_{\Sigma}|}.$
Putting (\ref{bkt}) in (\ref{vi}) and  assuming the adiabatic limit $|\vec{B}_{a}(\vec{k})|\gg \vec{B}_{\perp},$ we yield \beq v_{i} = -\gamma \frac{\partial|\vec{B}_{a}(\vec{k})|}{\partial p_{i}} - \frac{\hbar}{2}(\dot{\vec{n}}_{a}\times \vec{n}_{a}).\frac{\partial\vec{n}_{a}}{\partial p_{i}}.\label{cru}\eeq Writing $\dot{\vec{n}}_{a} = \dot{k}_{j}\frac{\partial\vec{n}_{a}}{\partial k_{j}}$ in (\ref{cru}) and rearranging the terms, we obtain the spin dependent velocity as
\beq v_{i} =  -\gamma \frac{\partial|\vec{B}_{a}(\vec{k})|}{\partial p_{i}} - \frac{\dot{\vec{k}}_{j}}{2}(\frac{\partial\vec{n}_{a}}{\partial k_{i}}\times \frac{\partial\vec{n}_{a}}{\partial k_{j}}).\vec{n}_{a} .\label{vii}\eeq The first term in the r.h.s of eqn. (\ref{vii}) gives the velocity component due to inhomogeneity in $\vec{B}_{a}(\vec{k}).$ The second term gives the anomalous part \cite{haldane} of the spin dependent velocity of electron which is related to the momentum space Berry curvature.  Noting the relation (\ref{omega}), the expression for the anomalous part of the spin dependent velocity in a linearly accelerated frame  can be written as
\beq v^{ano}_{i} = -\epsilon_{ijk}\dot{k}_{j}\Omega_{k}(\vec{k}),\label{vkl} \eeq
which can be compared with the  classic result of Sundaram and Niu \cite{niu}. In presence of $ \Omega(\vec{k})$ the equation of $\dot{\vec{r}}$  for a wave packet is \cite{niu}
\beq \dot{\vec{r}} = \frac{1}{\hbar}\frac{\partial\epsilon_{s}}{\partial\vec{k}} - \dot{\vec{k}}\times \Omega(\vec{k}).\label{rk}\eeq The anomalous velocity proportional to the Berry curvature is of great physical significance in the context of spin Hall effect.  The anomalous part of the  velocity actually represents the spin Lorentz force and is responsible for the spin Hall effect. In our formulation, the anomalous part of the velocity arises because of the presence of $\vec{B}_{\perp}$ which is effective due to the time dependent acceleration in the system. Surely, this Berry curvature in the momentum space is related to a Berry gauge field $A_{0}(t)$ and this time dependent gauge field  completes  the explanation of SHE.


\section{Conclusion}
In this paper we have studied theoretically the generation of spin current from the covariant Dirac equation in a linearly accelerating system in the presence of electromagnetic fields . We have derived  the expressions  for the inertial spin current and conductivity. The effect of acceleration is explicit in both the expressions  for the charge and spin component of the current whereas in the expressions for the conductivity the effect of acceleration is not manifested. For a constant external electric field, the spin current is found to vanish for a particular value of acceleration $a_c.$

We also discuss the characteristic  features of the momentum  space Berry curvature from the perspective of spin Hall effect for both the cases of time independent and time dependent acceleration. This in turn shows the  effect of acceleration  on the the time independent and time dependent effective magnetic fields present in the system. In particular, for time dependent acceleration, the expression for the spin polarization has been derived. We also demonstrate  the existence of an anomalous term present in the expression of the spin dependent velocity which represents a spin Lorentz force like term in the momentum space.

\vspace*{1cm}

{\bf{Acknowledgement}}: Authors appreciate discussions with Subir Ghosh.



\begin{thebibliography}{999}
\bibitem{wolf}S. A. Wolf, D.
D. Awschalom, R. A. Buhrman, J. M. Daughton, S. von
Molnar, M. L. Roukes, A.Y. Chtchelkanova, and D. M.
Treger, Science {\bf 294}, 1488 (2001).
\bibitem{zutic}I. Zutic, J. Fabian, and S. D. Sarma, Rev. Mod. Phys.
{\bf 76}, 323 (2004).
\bibitem{sh1} J. E. Hirsch, Phys. Rev. Lett. {\bf 83}, 1834 (1999)(arXiv:cond-mat/9906160).
\bibitem{spinh}  M. I. Dyakonov and V. I. Perel, Sov. Phys. JETP Lett. {\bf 13}: 467
(1971).
\bibitem{sh2}  Y. K. Kato, R. C. Myers, A. C. Gossard, and D. D. Awschalom, Science {\bf 306}, 1910 (2004);
J. Wunderlich, B. Kaestner, J. Sinova, T. Jungwirth, Phys. Rev. Lett. {\bf 94}, 047204 (2005);
S. O. Valenzuela and M. Tinkham, Nature {\bf 442}, 176 (2006).
\bibitem{29}D. Culcer, J. Sinova, N. A. Sinitsyn, T. Jungwirth and A.
H. MacDonald and Q. Niu, Phys. Rev. Lett. {\bf 93}, 046602 (2004); J. Shi, P. Zhang, D. Xiao and Q. Niu, Phys. Rev. Lett.{\bf 96} , 076604 (2006).
\bibitem{30} B. A. Bernevig, Phys. Rev. B {\bf 71}, 073201 (2005).
\bibitem{50}X.-J. Liu, X. Liu, L. C. Kwek, C. H. Oh, Phys. Rev. Lett.
{\bf 98},  026602 (2007).
\bibitem{51} J. Shibata, H. Kohno, Phys. Rev. Lett. {\bf 102},
086603 (2009).
\bibitem{52} M. Gradhand, D. V. Fedorov, P. Zahn, I. Mertig, Phys.
Rev. Lett. {\bf 104},  186403 (2010).
\bibitem{54}V. Ya. Kravchenko, Phys. Rev. Lett. {\bf 100}, 199703 (2008);
E. M. Chudnovsky, Phys. Rev. Lett. {\bf 100}, 199704 (2008);
\bibitem{56}S. Murakami, N. Nagaosa, S.-C. Zhang, Science, {\bf 301},
 1348 (2003);
\bibitem{58}S. Murakami, N. Nagaosa, S.-C. Zhang, Phys. Rev. B {\bf 69},
 235206 (2004).
\bibitem{60} J. Sinova, D. Culcer, Q. Niu, N. A. Sinitsyn, T. Jung-
wirth, A. H. MacDonald, Phys. Rev. Lett. {\bf 92},
126603 (2004).
\bibitem{barnett}S. J. Barnett, Phys. Rev. {\bf 6}, 239 (1915).
\bibitem{ein}A. Einstein and W. J. de Haas, Verh. Dtsch. Phys. Ges.
{\bf 17}, 152 (1915).
\bibitem{tol} R. T. Tolman and T. Stewart, Phys. Rev. {\bf 8},  97 (1916).
\bibitem{o} F. W. Hehl and Wei Tou Ni, Phys. Rev. D {\bf 42} (1990).
\bibitem{matsuoprl}M. Matsuo, J. Ieda, E. Saitoh, S. Maekawa ,Phys. Rev. Lett {\bf106}, 076601 (2011).
\bibitem{c}M. Matsuo et.al., Physical Review B  {\bf 84}, 104410 (2011).
\bibitem{berry}M.V. Berry, Proc. R. Soc. London A {\bf 392}, 45 (1984) .
\bibitem{xiaorev}D. Xiao, M.Chang and Q. Niu, Rev. Mod. Phys.{\bf 82}, 1959 (2010).
\bibitem{bpb} B. Basu, S. Dhar and P. Bandyopadhyay, Phys. Rev.B {\bf 69}, 094505 (2004).
\bibitem{carolo} A.C. M. Carollo and J. Pachos, Phys. Rev. Lett {\bf 95} 157203 (2005).
\bibitem{zhu}S. L. Zhu, Phys. Rev. Lett {\bf 96} 077206 (2006).
\bibitem{bb}B. Basu,  Europhys. Lett., {\bf 73}, 833 (2006).
\bibitem{pra2}B. Basu, P. Bandyopadhyay, and P. Majumdar, Phys. Rev. A 86, 022303 (2012).
\bibitem{shen}S. Q. Shen, Phys. Rev. B {\bf 70}, 081311(R)(2004).
\bibitem{46}A. Berard, H. Mohrbach, Phys. Lett. A {\bf 352}, 190 (2006).
\bibitem{78}B. Basu and P. Bandyopadhyay, Phys. Lett. A {\bf 373}, 148 (2008).
\bibitem{sg} S. Dhar, B. Basu and Subir Ghosh, Phys. Lett. A {\bf 371} 406 (2007).
\bibitem{horvathy}C. Duval, Z. Horvath and P. Horvathy, J. Gem. Phys. {\bf 57}, 925 (2007).
\bibitem{bliokh}K. Y. Bliokh, J. Opt. A: Pure Appl. Opt. {\bf 11}, 094009 (2009).
\bibitem{graph1} D. Xiao, W. Yao and Q. Niu, Phys. Rev. Lett. {\bf 99}, 236809 (2007).
\bibitem{graph2} P. Goselin, A. Berard and H. Mohrbach, Eur. Phys. J. C. {\bf 59}, 883 (2009).
\bibitem{n}E. M. Chudnovsky, Phys. Rev. Lett.{\bf 99}, 206601 (2007).
\bibitem{gre} W. Greiner, {\it{Relativistic Quantum Mechanics:Wave Equation}} (Springer-Verlag, Berlin, 2000), p.277
\bibitem{m} L.L Foldy and S.Wouthuysen, Phys Rev {\bf 78}, 29 (1950).
\bibitem{winkler} R. Winkler, {\it{Spin-orbit Coupling Effects in Two-Dimentional Electron and Hole Systems}} (Springer-Verlag, Berlin, 2003)
\bibitem{fujita}T Fujita, M B A Jalil and S G Tan, New Journal of Physics {\bf 12}, 013016 (2010).
\bibitem{rashba}E.I Rashba Sov. Phys. Solid State {\bf 2}, 1224 (1960); Y. Bychkov and E.I Rashba JETP Lett. {\bf 39}, 78 (1984).
\bibitem{haldane}F. D. M. Haldane, Phys. Rev. Lett {\bf 93}, 20 (2004).
\bibitem{niu}G. Sundaram and Q. Niu, Phys. Rev. B {\bf 59}, 14915 (1999).
\end{thebibliography}
\end{document}